\documentclass[sigconf]{acmart}

\usepackage{svg}
\usepackage{subcaption}
\usepackage{xcolor,soul}
 \sethlcolor{lightgray}
\usepackage{ftnxtra}

\AtBeginDocument{%
  \providecommand\BibTeX{{%
    \normalfont B\kern-0.5em{\scshape i\kern-0.25em b}\kern-0.8em\TeX}}}

\setcopyright{acmcopyright}
\copyrightyear{2022}
\acmYear{2022}
\acmDOI{XXXXXXX.XXXXXXX}

\acmConference['MM 22']{the 30th ACM International Conference on Multimedia October 10--14, 2022, Lisbon, Portugal}{October 10--14, 2022}{Lisbon, Portugal}
\acmPrice{15.00}
\acmISBN{978-1-4503-XXXX-X/18/06}

\begin{document}

\title{End-to-End and Self-Supervised Learning for ComParE 2022 Stuttering Sub-Challenge}

 \author{Shakeel A.~Sheikh$^1$, Md Sahidullah$^1$, Fabrice Hirsch$^2$, Slim Ouni$^1$}

\affiliation{\institution{Universite de Lorraine, CNRS, Inria, LORIA, F-54000, Nancy, France$^1$ \\  Universite Paul-Valery Montpellier, CNRS, Praxiling, Montpellier, France$^2$}
  \country{}
}


\renewcommand{\shortauthors}{Shakeel A.~ Sheikh, et al.}

\begin{abstract}
   In this paper, we present end-to-end and speech embedding based systems trained in a self-supervised fashion to participate in the ACM Multimedia 2022 ComParE Challenge, specifically the stuttering sub-challenge. In particular, we exploit the embeddings from the pre-trained Wav2Vec2.0 model for stuttering detection (SD) on the KSoF dataset. After embedding extraction, we benchmark with several methods for SD. Our proposed self-supervised based SD system achieves a UAR of 36.9\% and 41.0\% on validation and test sets respectively, which is 31.32\% (validation set) and 1.49\% (test set) higher than the best (DeepSpectrum) challenge baseline (CBL). Moreover, we show that concatenating layer embeddings with Mel-frequency cepstral coefficients (MFCCs) features further improves the UAR of 33.81\% and 5.45\% on validation and test sets respectively over the CBL. Finally, we demonstrate that the summing information across all the layers of Wav2Vec2.0 surpasses the CBL by a relative margin of 45.91\% and 5.69\% on validation and test sets respectively. 
   
\end{abstract}




\maketitle

\section{Introduction}

Stuttering is a speech impairment that impairs a person's ability to communicate and is a neuro-developmental speech disorder mostly characterized by involuntarily blocks/stop gaps, repetitions, prolongations, and interjections~\cite{duffy2019motor, ward2017stuttering}. These uncontrolled utterances are usually accompanied by psychological and linguistic variabilities~\cite{kehoe2006speech, guitar2013stuttering}. In addition, unique behaviors such as head shaking, lip tremors, fast eye blinks, and unusual lip shapes are commonly associated with these uncontrolled utterances~\cite{guitar2013stuttering}. These unique abnormal behaviors make it difficult for people who stutter (PWS) to communicate properly and, as a result, have a detrimental impact on their lives~\cite{nsastutter}.
Speech therapy is frequently used by PWS to cope with their problem, where stuttering is identified using a variety of hearing and brain scan tests~\cite{sheikh2021machine, Ingham1996FunctionallesionIO, smith2017stuttering}, however, such manual methods are extremely arduous, time-consuming, and expensive. Stuttering detection (SD) can also be adapted towards voice assistants such as Cortona, Alexa, etc. where the automatic speech recognition systems fail to recognize stuttered speech~\cite{asrstuttering}.

\par 
Most of the previous work in SD explored traditional classifiers mainly on hand-engineered features such as Mel-frequency cepstral coefficients (MFCCs), etc~\cite{sheikh2021machine}. However, the trend has recently shifted towards the deep learning paradigm-based SD. \citet{tedd} 
exploited residual network in conjunction with bi-directional long short-term memory (ResNet+BiLSTM) on a small subset of speakers (25) from the UCLASS dataset. \citet{stutternet} approached SD as a multi-class classification problem and introduced \emph{StutterNet} which is based on a time delay neural network (TDNN) on a large set of speakers (100+) on the UCLASS dataset. Another study by \citet{sep28k} proposed a new SEP-28k stuttering dataset and used ConvLSTM on top of phoneme features for SD. A recent study carried out by~\citet{stutternetmtl} introduces adversarial learning to unlearn the podcast information from a speech utterance to learn robust stutter representations that are stutter discriminative, and at the same time are podcast invariant. 

Recently, \citet{bayerw2v2} demonstrated the usefulness of leveraging features extracted from self-supervised pre-trained models that are trained on enormous datasets by utilizing the Wav2Vec2.0 model as a feature extractor.
In a related study by \citet{sheikhw2v2}, embeddings from the emphasized channel attention, propagation, and aggregation-time-delay neural network (ECAPA-TDNN) were also examined in addition to Wav2Vec2.0 embeddings in SD domain.

\par 
Our contribution to ComParE 2022 KSF-C stuttering sub-challenge explores end-to-end and self-supervised systems. For end-to-end systems, we use \emph{StutterNet}~\cite{stutternet} and ResNet+BiLSTM~\cite{tedd} with MFCC input features computed from KSoF dataset, and for addressing limited data issue, we explored self-supervised pre-trained speech embeddings extracted from various layers of Wav2vec2.0 model~\cite{wav2vec2}. In addition, we demonstrate the impact and concatenation of Wav2Vec2.0 layer embeddings in SD. 

\section{Overview of the challenge}
The ACM Multimedia 2022 Computational Paralinguistics ChallengE (ComParE) proposes four sub-competitions including \textit{Vocalisations}, \textit{Stuttering}, \textit{Activity}, \& \textit{Mosquitoes}~\cite{Schuller22-TI2}. Among these, we only focus on the \textit{Stuttering} (KSF-C) sub-challenge. In this sub-challenge, the task is to classify the speech segments into one of the eight categories including filler, garbage, prolongation, sound repetition, block, modified, word repetition, and no\_disfluency. The details about the stuttering challenge, KSoF dataset, and its partitioning are available in~\cite{Schuller22-TI2}.

\section{System description}

\subsection{End-to-End Systems}
In end-to-end speech classification systems, the model directly maps the input speech (e.g. raw speech or features such as MFCCs~\cite{huang2001spoken}) into its corresponding class. This involves a single-phase model training. In this work, we utilize two end-to-end models: \emph{StutterNet} and ResNet+BiLSTM~\cite{tedd}. 
\subsubsection{StutterNet} 
The \emph{StutterNet} is based on a time delay neural network which has been proven effective in capturing temporal and contextual aspects of speech signal~\cite{tdnn, peddinti15b_interspeech, stutternet}. It is composed of five time delay layers with the initial layers capturing smaller contexts and deeper ones learning wider contexts in contrast to the standard neural networks which learn wider contexts at initial layers. The output activations from the last layer are fed to the statistical pooling layer to compute the mean and standard deviation across the temporal dimension. This is followed by three fully connected layers with each layer followed by a ReLU activation function. We apply batch normalization after each layer except the statistical pooling layer.

\par 
The KSoF dataset provided in this challenge is highly imbalanced~\cite{Schuller22-TI2}. To address the class imbalance, we used cost-level and architecture-level approaches. For cost-level, we penalize the majority class by modifying the standard cross entropy as:
\begin{equation}
     \mathbf{L}_{\mathrm{WCE}} = \frac{1}{\mathbf{B}}\sum\limits_{b=1}^{\mathbf{B}}\frac{\sum\limits_{i}^{N} \alpha_i * \log(p_i)}{\sum\limits_{i,~i \in \mathbf{B}}^{N} \alpha_i}
     \label{eq:wce}
\end{equation}
where $\mathbf{B}$ is total batches, N is the number of stuttered speech samples in a batch $b_i$, $\alpha_i = \frac{\mathcal{N}}{C * \mathcal{N}_{i}}$ ($\mathcal{N}$ is the number of training samples, $C$ is number of classes, $\mathcal{N}_i$ is the number of training samples for class $i$), $p_i = \Big( \frac{e^{c_i}}{\sum_{j=1}^C e^{c_j}} \Big)$ is the predicted probability of class $c_i$ of sample $i$.

\par 
For architecture-level, we use multi-branch training scheme similar to the work from \citet{sep28k, stutternetmtl, sheikhw2v2}. This comprises a base encoder $\mathbf{E}$ ($\theta_{e}$) followed by two parallel branches referred as \emph{DisfluentBranch} $\mathcal{D}$ ($\theta_{d}$) and \emph{FluentBranch} $\mathcal{F}$ ($\theta_{f}$).
The embeddings generated by  $\mathbf{E}$ ($\theta_{e}$) are simultaneously passed to both the \emph{FluentBranch} and \emph{DisfluentBranch}, where the \emph{FluentBranch} is trained to distinguish between fluent and disfluent samples, and the \emph{DisfluentBranch} is trained to differentiate and classify the disfluent\footnote{If the prediction of a sample in $\mathcal{F}$ is not fluent, then the predictions of $\mathcal{D}$ are taken into consideration to reveal the disfluent class category} sub categories with an overall objective function to optimize as:
\begin{equation}
    \mathcal{L}_{\mathrm{tot.}}(\theta_{e}, \theta_{f}, \theta_{d})  =  \mathcal{L}_f(\theta_{e}, \theta_{f}) + \mathcal{L}_d(\theta_{e}, \theta_{d})
    \label{eq:mbloss}
\end{equation}

\subsubsection{ResNet+BiLSTM}
The ResNet+BiLSTM based SD, proposed by \citet{tedd} comprises residual unit with 18 convolution blocks to capture stutter-specific features. These features are then provided to two recurrent layers, with each layer having 512 BiLSTM units. Moreover, we also employed a multi-branched version of ResNet+BiLSTM in a similar fashion applied to \emph{StutterNet}. 

\begin{figure}
     \centering
        \includegraphics[scale=0.4]{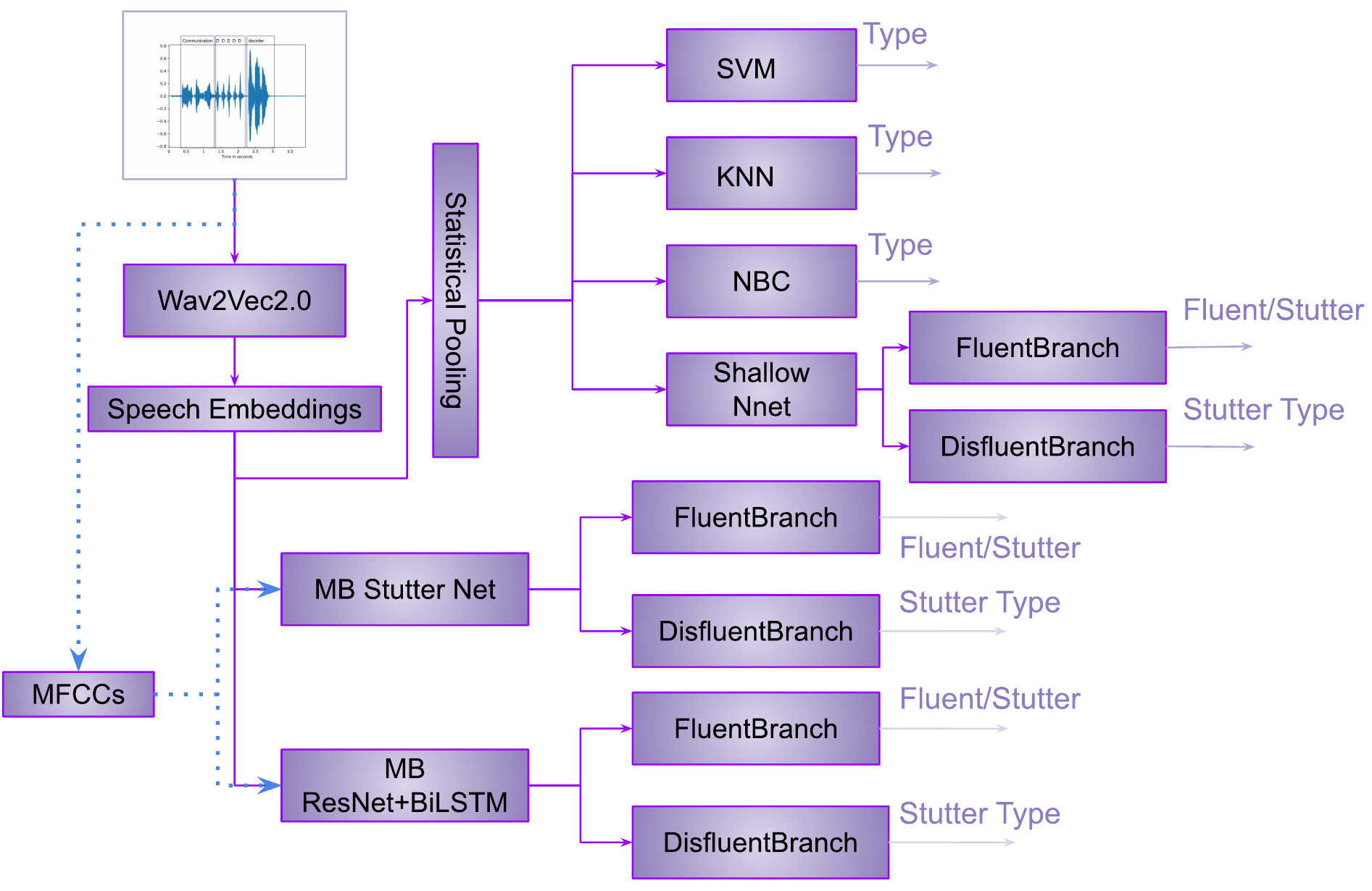}  \vspace{-0.3cm}

    \caption{\small Block Diagram of the Proposed Pipeline for SD.}  
    \label{fig:wa2vec2.0}   \vspace{-0.45cm}
\end{figure}
\subsection{Self-Supervised Framework}
Due to the availability of small-sized stuttering datasets, the deep learning paradigm hasn't shown much improvement in SD in comparison to other speech domains such as automatic speech recognition~\cite{wav2vec2}, speaker verification~\cite{BAI202165}, emotion detection~\cite{speechemotion, edwav2vec2}, speaker diarizarion~\cite{spkdiarize}, etc. The small datasets are limited in capturing accents, linguistic content, age group, different speaking styles, etc. Moreover, the speech pathology datasets are very expensive to collect, as a result, this prevents the adoption of advanced deep learning models for SD. To overcome this bottleneck, we employ self-supervised learning (SSL), where we extract the features from the model pre-trained on a huge dataset for downstream SD tasks.
SSL makes use of the data's underlying structure. In SSL classification systems, the model is first pre-trained on some pre-auxiliary task to capture rich embeddings from the innate structure of the data~\cite{schneider19_interspeech, wav2vec2, mohamed2022self, ericsson2022self}. These embeddings are then used for other downstream classification tasks. In this paper, we use speech representations from various layers of the pre-trained Wav2Vec2.0 model for the downstream SD task.

\subsubsection{Wav2Vec2.0}
The Wav2Vec2.0 model is a three-module-based self-supervised representation learning framework for raw audio, pre-trained on LibriSpeech dataset followed by fine-tuning towards automatic speech recognition task using connectionist temporal classification loss function. The three modules consist of feature encoder $\mathbb{F}$, contextual block $\mathbb{C}$ and quantization block $\mathbb{Q}$. The $\mathbb{F}$ encodes the raw input signal $\mathcal{X}$ into local features. These encoded features are then passed to $\mathbb{C}$ and $\mathbb{Q}$ modules to learn contextual speech representations.
 Several pre-trained models of contextual embedding dimensions of 768 (base) and 1024 (large) have been released. The model is trained in a self-supervised manner to learn the speech representations of an utterance by optimizing contrastive loss function using equation~(\ref{eq:contrastiveloss}) as below:
 \begin{equation}
    \mathcal{L}_{c} = -\log\frac{\exp(sim(c_t, q_t)/\tau)}{\sum_{\tilde{q}\in Q_t}\exp(sim(c_t, \tilde{q})/\tau)}
    \label{eq:contrastiveloss}
\end{equation}
where $sim(c_t, q_t) = c_{t}^{T}q_t/||c_t||~||q_t||$ is the cosine similarity between the contextualized transformer vector $c_t$ and  quantized vector $q_t$.
 Equation~(\ref{eq:contrastiveloss}) is then augmented by a diversity loss. For details, please refer to the paper by \citet{wav2vec2}.

 \par 
 In this paper, we use only the base pre-trained model of 768 embedding dimensions (trained on 960 hours of data) and we extract embeddings from the $\mathbb{F}$  block and from 12 layers of $\mathbb{C}$ block for SD. Moreover, we use statistical pooling over the temporal domain and concatenated the mean and standard deviation, resulting in a $768\times2$-dimensional feature vector before passing it to the downstream classifiers except multi-branched (MB) \emph{StutterNet}, where we pass the speech embeddings directly.

\par

\section{Results and Discussion}
\subsection{Training Details}
We train the MB \emph{StutterNet} and a shallow neural network (NNet) with Adam optimizer on a  cross entropy loss function ($L = Lf + L_d$, $L_f$:\emph{FluentBranch loss}, $L_d$:\emph{DisfluentBranch loss} similar to equation~(\ref{eq:mbloss})) using a learning rate of 1e-2, with a batch size of 128 over 50 epochs. The shallow NNet is composed of two branches with \emph{FluentBranch} and \emph{DisfluentBranch} with each branch having three fully connected (FC) layers. Each FC layer is followed by a ReLU activation function~\cite{activationssurvey} and a 1D-batch normalization~\cite{batchnorm}. A dropout~\cite{JMLR:v15:srivastava14a} of 0.3 is applied to the first two FC layers. A patience of seven is applied on a validation loss to stop the training.
We train the MB \emph{StutterNet}  with two input features including $20\times T$ MFCCs and $768\times T$ speech embeddings extracted from Wav2Vec2.0.
For support vector machines (SVM)~\cite{murphy2012machine}, we experiment with different kernels including linear, polynomial, Radial basis function, and sigmoid, however, in this paper, we report the results only with the linear kernel. For K-nearest neighbor (KNN)~\cite{murphy2012machine}, we choose the value of K = 5 empirically by grid search using the elbow method with $p=2$ (Euclidean) distance metric. A statistical pooling (mean and standard deviation) is applied to Wav2Vec2.0 $768\times T$ speech embeddings before passing them to KNN, naive Bayes classifier (NBC)~\cite{murphy2012machine}, SVM, and NNet downstream classifiers. We use PyTorch~\cite{pytorch} and Scikit-learn~\cite{scikit-learn} for implementation purposes. 
\par

\begin{figure}
     \centering
     \includegraphics[scale=0.5]{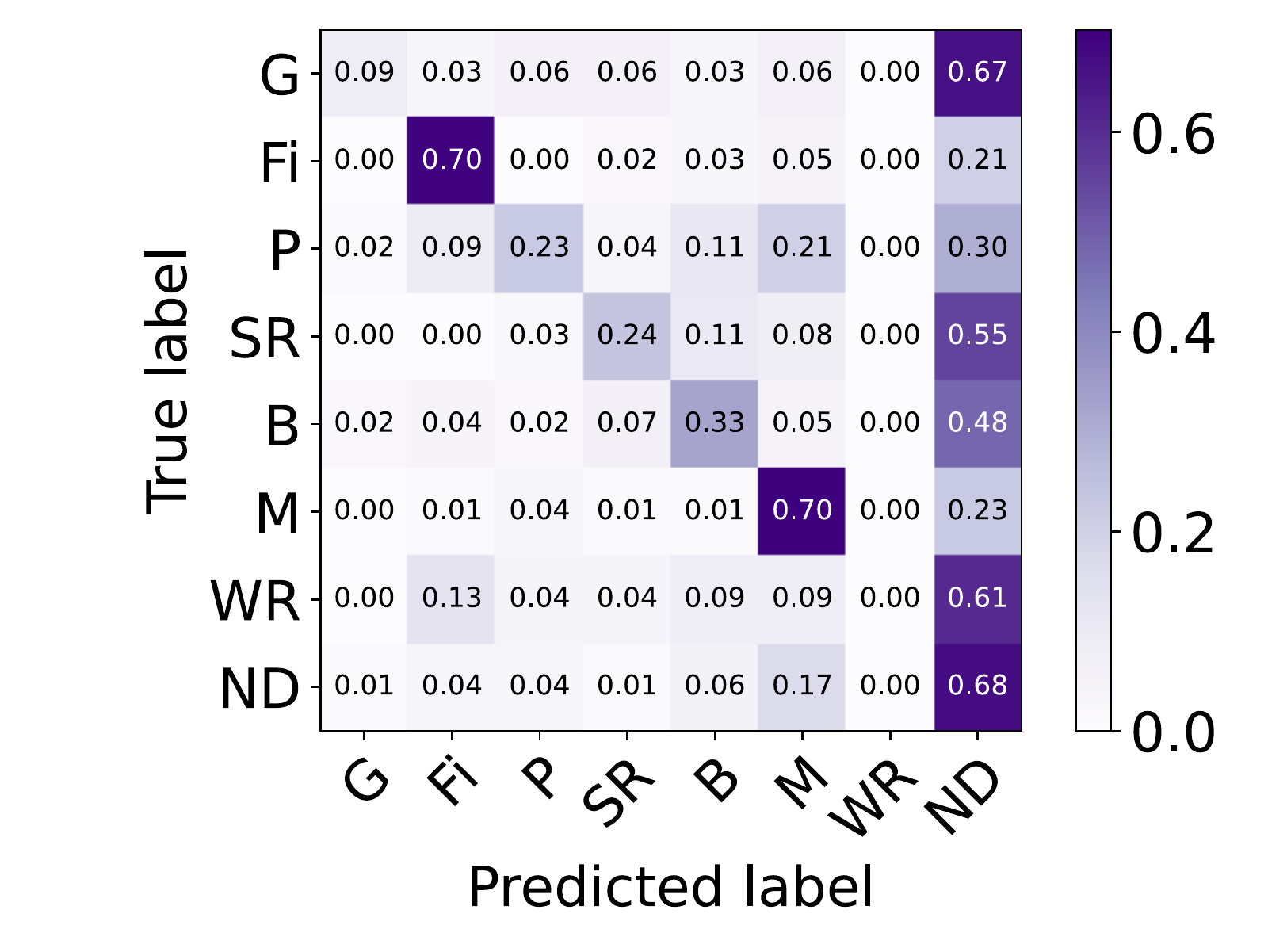} \vspace{-0.5cm}
    \caption{\small Confusion Matrix of SVM on Val. Set with L5 Embeddings\footnotemark}   
    \label{fig:cmatrix}  \vspace{-0.5cm}
\end{figure}
\footnotetext{Please refer to Table.~\ref{tab:results} for class names.}

\begin{table*}
  \caption{\small SD Results in Accuracy and UAR (G: Garbage, Fi: Fillers, P; Prolongation, SR: SoundRepetition, B: Block, M: Modified, WR: Word Repetition, ND: no\_disfluencies, TA; Total Accuracy, BL: Baseline, CE: Cross Entropy, WCE: Weighted Cross Entropy, UAR: Unweighted Average Recall, $L_i$: Embeddings from Layer $i$, ":" is Concatenation, $\sum_i L_i$: Summing Information Over all Embedding Layers).}
  \label{tab:results}
 \scalebox{0.8}{ \begin{tabular}{ccccccccccccc}
    \toprule
    Model &Feature&G&Fi&P&SR&B&M&WR&ND&TA&UAR\%(Val)&UAR\%(Test)\\
    \midrule
    
    ResNetBiLSTM+CE (BL)&MFCCs&	0.00&	0.00&	0.00	&7.89	&38.46	&60.54	&0.00	&58.33	&42.16&	20.7&NA\\
    
   ResNetBiLSTM+WCE&MFCCs&	 42.42&	0.00&	60.38&	0.00&	01.92&	18.38&	8.70&	01.13&	09.06&	16.6 & NA\\
MB ResNetBiLSTM &L5 &0.00&	0.00&	35.85&	7.89&	64.42&	65.95&	0.00&	16.67& 	29.02&	23.8&NA\\


   SB \emph{StutterNet} + CE (BL) &MFCCs& 03.03&	03.92&	09.43&	0.00&	41.35&	71.35&	0.00&	52.70&	42.67&	23.0&NA\\
   SB \emph{StutterNet} + WCE &MFCCs& 51.52& 20.59&	26.42&	10.53&	19.23&	43.24&	17.39&	03.60&17.92&	24.0&NA\\
  MB \emph{StutterNet}&MFCCs& 21.21&	14.71&	35.85&	18.42&	43.27&	78.38&	0.00&	20.27&	33.40&	29.0 & NA\\

   \midrule
    SVM (Linear)&L5&9.09&	69.61&	22.64&	23.68&	32.69&	70.27&	0.00&	67.57&	56.93&	36.9&41.0\\
     SVM (Linear)&MFCCs:L5&21.21&	59.80&	26.42&	26.32&	39.42&	71.35&	0.00&	56.08&	52.34&	37.6&42.6\\
NBC & L9&45.46&	41.18&	30.19&	26.32&	39.42&	70.81&	0.00&	08.56&	29.84&	32.7&NA\footnotemark	\\
NNet & L5& 6.06&	59.80&	26.42&	10.53&	0.00&	75.14&	4.35&	\textbf{81.08}&	\textbf{59.17}&	32.9& NA\\

MB \emph{StutterNet}&L5&15.15&71.57&	32.08&	15.79&	39.42&	76.22&	0.00&	57.43	&	54.79&	38.5&40.3\\
MB \emph{StutterNet}&L5:L9 &03.03&	81.37&	28.30&	\textbf{28.95}&	50.96&	74.60&	4.35&	54.50&		55.40&	40.8&42.6\\

MB \emph{StutterNet}&$\sum_iL_i$&12.12&\textbf{78.43}&\textbf{39.62}&	26.32&	58.65&	\textbf{80.00}&	0.00&	32.88&	47.86&	\hl{\textbf{41.0}}&\hl{\textbf{42.7}}\\

  \bottomrule
\end{tabular}}
\end{table*}

 \begin{figure}
    

\includegraphics[scale=0.2]{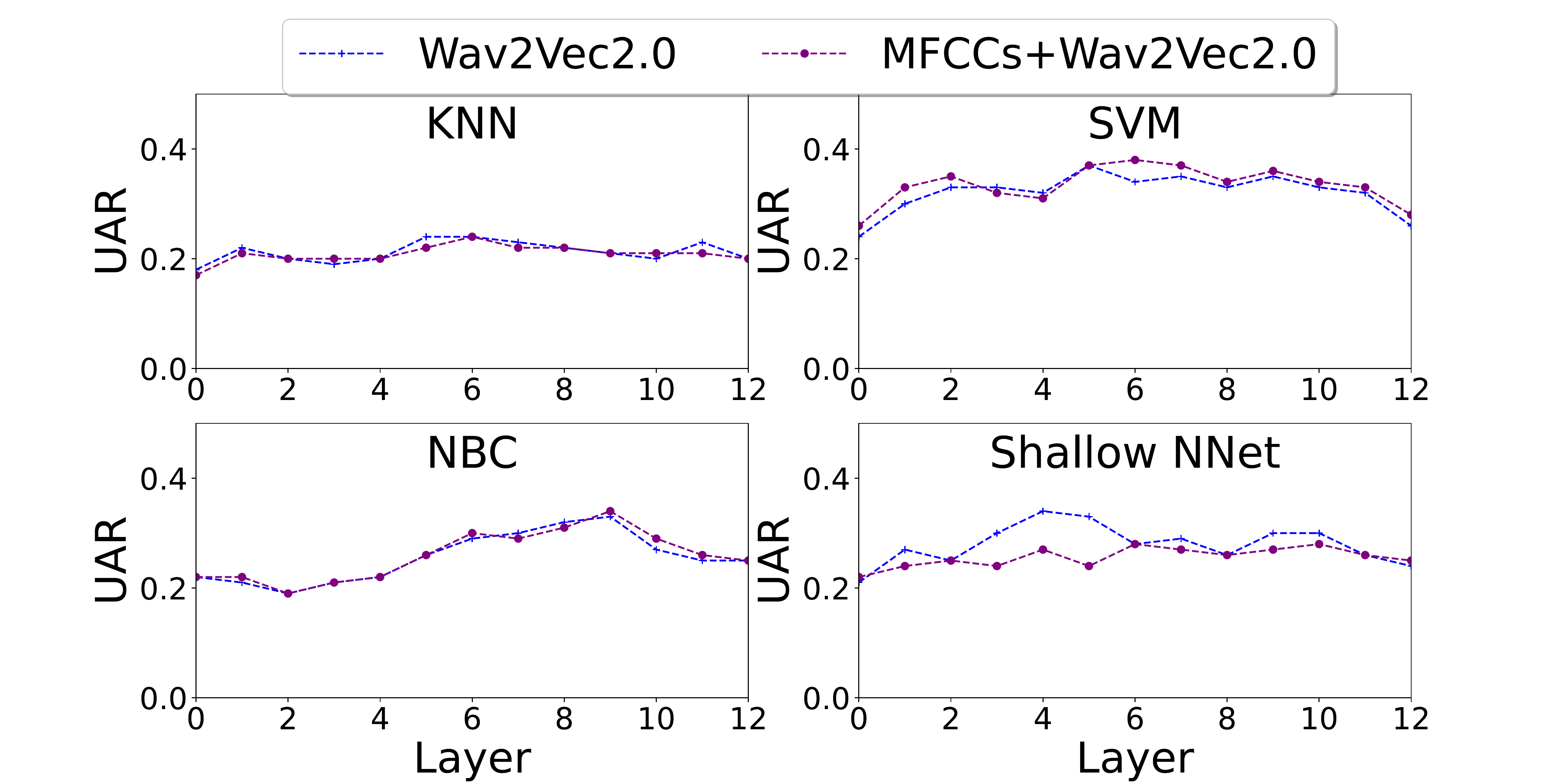}
\vspace{-0.6cm}
 \caption{\small Comparison of Various Wav2Vec2.0 Contextual Layers and its Concatenation with MFCCs in SD with KNN, NBC, SVM, and Shallow NNet Classifiers where Y-axis Represents Unweighted Average Recall, and X-axis Represents Wav2Vec2.0 Layer Embeddings.}  \vspace{-0.2cm}
 \label{fig:layerimpact}
\end{figure} 

\subsection{Evaluation}
Table~\ref{tab:results} summarizes the results of our experiments on KSoF dataset provided in this ComParE challenge~\cite{Schuller22-TI2}. We first train the end-to-end single branched version of \emph{StutterNet} and ResNet+BiLSTM~\cite{tedd} models on MFCC input features using a standard cross entropy loss function. Due to the class imbalance nature in KSoF dataset provided in this ComParE challenge, the models are skewed towards majority class recognition including \textit{blocks}, \textit{modified}, and \textit{no\_disfluencies} as can be confirmed by the results Table~\ref{tab:results}. We then applied cost-based approach by modifying the standard cross entropy using equation~(\ref{eq:wce}) with an aim to focus more on the minority classes. We found a huge improvement in garbage and prolongation categories using ResNet+BiLSTM, and, we found \textit{word repetitions}, \textit{fillers}, and \textit{sound repetitions} are the most difficult to recognize for ResNet+BiLSTM based model. For WCE based \emph{StutterNet}, we found improvement in \textit{garbage, fillers, prolongations and word repetitions}. Similar to ResNet+BiLSTM, we found that the \textit{sound repetitions} are still the most challenging to identify. Further, we applied a multi-branched training scheme to our baselines, and, we found that except for \textit{word repetition and no\_disfluencies} classes, the MB \emph{StutterNet} shows respectively a huge improvement of 600\%, 275\%, 280\%, 4.63\%, and 9.85\% in \textit{garbage, fillers, prolongation, sound repetition, blocks, and modified} over the baseline. 
\par 
In the SSL setup, we experiment with different speech embeddings extracted from various layers of a Wav2Vec2.0 pre-trained model. The extracted speech embeddings improve the detection performance of almost all the classes over the baseline. Similar to \citet{Schuller22-TI2}, we also found that the \textit{word repetitions} are the most challenging ones to recognize. Among the various layers, layer five (L5) shows the best performance on UAR with all the downstream classifiers except for NBC, where, speech embeddings from layer nine (L9) perform better. Figure~\ref{fig:layerimpact}. shows the impact of various speech layer embeddings in SD with downstream classifiers. As presented by the plots, the first initial and last layers show lower UAR in comparison to the middle layers. We hypothesize that the lower layer speech representations contain information only from smaller contexts, and passing this information after statistical pooling to SVM, KNN, NBC, and NNet further inhibits it in learning stutter-specific patterns. Furthermore, \citet{shah2021all} have also shown that the middle layers are good at capturing fluency and pronunciation features which are extremely important for stuttering detection, while initial layers of Wav2Vec2.0 exhibit more deviation in learning fluency (e.g., speech rate, pauses, etc.) and pronunciation (e.g., vowels, consonants, syllables, stress, etc.) features. The trend also shows that the UAR curve decreases for the last layers. This is most likely due to the reason that the Wav2Vec2.0 model is fine-tuned and adapted towards the ASR task, and it is quite possible that the information such as prosody, stress, and emotion state gets diluted, as also can be confirmed from the study by \citet{shah2021all}. We also observe that the improvement gap between MFCC-based systems and Wav2Vec2.0 embedding-based systems is lower than the study carried out by \citet{sheikhw2v2}. The most obvious reason seems the linguistic information between the two. \citet{sheikhw2v2} used the SEP-28k corpus for the downstream SD task, which is also an English language stuttering dataset that coincides with the language of the LibriSpeech on which the Wav2Vec2.0 model was trained. Contrary to the SEP-28k dataset, there is a possibility of the loss of linguistic information while extracting embeddings of the KSoF German dataset (provided in this ComParE challenge) from the English-based Wav2Vec2.0 pre-trained model.

In addition, we experiment by concatenating the MFCC features to each speech embeddings layer of Wav2Vec2.0 after applying statistical pooling to both the features and using the downstream classifiers KNN, SVM, NBC, and Shallow NNet for SD, the plots of which are shown in Figure~\ref{fig:layerimpact}. We observe from the curves that the concatenation of MFCCs with layer embeddings only helps for the SVM classifier while for KNN and NBC, it remains almost unchanged. For shallow NNet, the UAR degrades while concatenating MFCC and speech embeddings. Over the CBL baseline, the SVM with speech embeddings and concatenated (MFCCs+speech embeddings) results in a relative improvement of 31.32\%, and 33.81\% on the validation set, and 1.49\% \footnotetext[4]{We are only allowed to submit five UAR results on the test set due to the restriction in this ComParE challenge.} and 5.45\% on the test set respectively. We make use of the fact that each layer of the Wav2Vec2.0 model has different information and passing the summation over all layers ($\sum_i L_i$) to MB \emph{StutterNet} results in a relative improvement of 45.91\% on the validation set and 5.69\% on the test set in UAR over the baseline (CBL) as shown in Table~\ref{tab:compare}. 
\par 

After analyzing various confusion matrices from the SD systems mentioned in Table~\ref{tab:results}, we found the trend that the most of the disfluent classes are still being falsely classified as ND (\textit{no\_disfluencies} class), and confusion in \textit{garbage} is maximum followed by \textit{word repetitions}, \textit{sound repetitions and blocks}. This makes intuitive sense because the \textit{repetitions} contain certain words or phrasal parts that, when examined individually, reveal themselves to be fluent utterances. Consider the phrase, "\textit{for for} the movement". The word \textit{for} is repeated twice, however, it is a fluent component if two \textit{fors} are examined separately. For \textit{garbage}, it is hard to understand the reason since it is either unintelligible or contains no speech. Furthermore, we observe that switching from inputting speech embeddings from a single layer to summing information across all layers ($\sum_i L_i$) reduces the confusion rate of various classes for MB \emph{StutterNet}.

\begin{table}
  \caption{\small UARs of the challenge baselines and our systems. (Val: Validation Set, Test: Test Set, CBL: Challenge Baseline)}
  \label{tab:compare}
 
  \scalebox{0.8}{\begin{tabular}{ccc}
   \toprule
  Approach & Val(\%) & Test(\%) \\
    \toprule
     ComParE~\cite{Schuller22-TI2} &30.2 &37.6\\
     auDeep~\cite{Schuller22-TI2} &17.7 &25.9\\
     BoAWs~\cite{Schuller22-TI2} &26.7 &32.1\\
     DeepSpectrum (CBL)~\cite{Schuller22-TI2} &28.1  &40.4  \\
     Fusion~\cite{Schuller22-TI2} &28.7 &38.3 \\
     \midrule
     (Ours) (SVM+L5) &36.9 & 41.0 \\
     (\textbf{Ours}) SVM+MFCCs:L5 &37.6&42.6\\
     (\textbf{Ours}$)~MB~\emph{StutterNet} + \sum_iL_i$ &\hl{\textbf{41.0}}&\hl{\textbf{42.7}}\\
      \bottomrule
      \vspace{-1.5cm}
\end{tabular}}
 
\end{table}

\section{Conclusion}

This work presents our contribution to the ComParE competition by addressing the stuttering sub-challenge using end-to-end and pre-trained self-supervised models. In comparison to the best baseline (CBL), the Wav2Vec2.0 speech embeddings based SD system outperforms by a relative margin of 5.69\% in UAR on the test set. The KSoF stuttering dataset provided in this challenge is highly imbalanced similar to the SEP-28k dataset, and in addition, the meta-data information such as age, transcription, and speaker information is also absent that could have been exploited to make the SD systems more robust.

\begin{acks}
This work was made with the support of the French National Research Agency, in the framework of the project ANR BENEPHIDIRE (18-CE36- 0008-03)
\end{acks}

\bibliographystyle{ACM-Reference-Format}
\bibliography{sample-base}

\end{document}